\documentclass[journal]{IEEEtran}
\ifCLASSINFOpdf
\else
\fi
\usepackage{subfigure}
\usepackage{graphicx}
\usepackage{epstopdf}
\usepackage{graphics}
\usepackage{moreverb}
\usepackage{algorithm}
\usepackage{amssymb}
\usepackage{amsmath,amscd,amssymb,amsthm}
\usepackage{fancyhdr}
\usepackage{algorithm}
\usepackage{etoolbox}

\begin{document}
%
\title{Multi-target Joint Detection, Tracking and Classification Based on Generalized Bayesian Risk using Radar and ESM sensors}
%
%
%

\author{Minzhe Li,
        Zhongliang Jing \\
        School of Aeronautics and Astronautics, Shanghai Jiao Tong University 
}

%
%

\markboth{ }%
{Shell \MakeLowercase{\textit{et al.}}: Bare Demo of IEEEtran.cls for IEEE Journals}
%



\maketitle



%
\IEEEpeerreviewmaketitle

\begin{abstract}
In this paper, a novel approach is proposed for multi-target joint detection, tracking and classification based on the labeled random finite set and generalized Bayesian risk using Radar and ESM sensors. A new Bayesian risk is defined for the labeled random finite set variables involving the costs of multi-target cardinality estimation (detection), state estimation (tracking) and classification. The inter-dependence of detection, tracking and classification is then utilized with the minimum Bayesian risk. Furthermore, the conditional labeled multi-Bernoulli filter is developed to calculate the estimates and costs for different hypotheses and decisions of target classes using attribute and dynamical measurements. Moreover, the performance is analyzed. The effectiveness and superiority of the proposed approach are verified using numerical simulations. 
\end{abstract}



\section{Introduction}
\indent \IEEEPARstart{M}ulti-target joint detection, tracking and classification (JDTC) using Radar and ESM sensors is a critical problem in airborne surveillance systems. In this problem, both kinematic measurements and attribute measurements are used to estimate the number of the targets, estimate their kinematic states, and determine their classes. Actually, these three subproblems are usually coupled: tracking may provide flight envelop and kinematic feature to distinguish the target type, according to the target class, appropriate dynamic models can be chosen for accurate tracking, and the change of the target number implies a modification of tracking and classification procedures \cite{R01}. Actually, multi-target JDTC is a joint decision and estimation (JDE) problem. \\
\indent Most traditional multi-target JDTC algorithms can be classified into the following categories. 
1) Estimation-Then-Decision (ETD): In this category, target tracking is usually performed using data from kinematic sensors, and the classification is then derived based on the flight envelopes and kinematic estimates \cite{R02}-\cite{R04}. The drawback of this two-step strategy is that, the classification is significantly dependent on the estimates. As shown in \cite{R05}, the classification performance was deteriorated due to the inaccurate state estimates derived with the error data association.
2) Decision-Then-Estimation (DTE). In this category, the decision is made using data from identity or attribute sensors, and the estimates are then calculated based on the decisions made before \cite{R103}. The disadvantage of this strategy is that, the error of the decision is not considered. In \cite{R102}, the state estimates were calculated with classification-aided data association, however, the classification was done without regarding the quality of the estimation it would lead to. 
3) Based on the joint probability density: In this category, the target state and class are inferred by the joint state-class probability density function. 
In \cite{R01}\cite{R06}\cite{R07}, the class dependent multi-target density was calculated using the particle implementation of PHD/MeMBer filter \cite{R08}\cite{R09} with corresponding motion model set, and the probability of target class could then be inferred by the weights of particles in the cluster. However, in these methods, the state and class of each target were not explicitly obtained. Furthermore, the overall performance may not be necessarily good because the final joint decision and estimation goal was not directly reached \cite{R10}. \\ 
\indent In \cite{R11}, Li proposed a new approach for the problems involving inter-dependent decision and estimation based on a generalized Bayesian risk. In this method, the decision and estimation costs were converted to a unified measure using additional weight coefficients, and the optimal solution was derived to minimize the Bayesian risk. Because the inter-dependence between decision and estimation was considered, this method is inherently superior to the conventional approaches. 
In \cite{R12}\cite{R13} the recursive JDE (RJDE) algorithm was developed to fit the dynamic system and solve target JTC problem. Moreover, a joint performance metric (JPM) was proposed for evaluating the overall performance. 
In \cite{R10}, the conditional JDE (CJDE) algorithm was proposed based on a new Bayesian risk defined conditioned on data and used to solve the target JTC and JDT problems \cite{R14}\cite{R15}. Because the estimates and costs were directly calculated using corresponding measurements once the decision is made, the computation of the algorithm is simplified greatly. \\ 
\indent In this paper, a novel approach is developed for multi-target JDTC based on the generalized Bayesian risk using Radar and ESM sensors. A new Bayesian risk is defined based on the labeled RFS involving the costs of multi-target detection, tracking and classification, and the optimal solution is then derived to minimize this new risk. Given the class decision sets of multiple targets, the posterior state estimates and class probabilities are calculated using kinematic measurements and attribute measurements within the Bayes recursion. 
For the explicit expression of the multi-target posterior density involving the measurement-target-associations (MTA's), the RFS based estimation and classification costs are exact calculated and the optimal JDE solution is directly derived. 
The Gaussian mixture implementation of the proposed algorithm is also developed, and the performance of the approach is analyzed.
Simulations show that the proposed approach performs better than traditional methods. \\
\indent This paper is organized as follows: An introduction of LMB filter and CJDE approach is presented in Section 2. The recursive multi-target JDTC algorithm is developed in Section 3. The simulation results of the proposed algorithm are provided in Section 4. Conclusions are summarized in Section 5. 

\section{Background}
\subsection{Labeled multi-Bernoulli RFS and multi-target Bayes filter}
\indent In \cite{R08}, a Bernoulli RFS was used to represents the uncertainty about the existence of a single object. The probability density function of a Bernoulli RFS $X$ can be given by
\begin{equation}
f(X)=\left\{\begin{aligned}
1-p, \qquad &if X=\emptyset \\
p\cdot f(x), \quad &if X=\{x\}
\end{aligned}
\right.
\end{equation}
As expressed in the equation, the Bernoulli RFS can either be empty with a probability of $1-p$, or have one element $x$ with probability $p$, and $f(x)$ is the probability density function of variable $x$ over space $\mathcal{X}$. \\
\indent In \cite{R17}, Vo et al. introduced the notion of labeled RFS. Assume that ${X}$ denotes the RFS of target states, the multi-target exponential of a real valued function $h$ for all the state vectors $x$ is $h^{X}\triangleq \prod_{x\in X}h(x)$, where $h^\emptyset = 1$. The Kronecker delta function and the inclusion function are
\begin{align}
\delta_{Y}({X})=\left\{
\begin{aligned}
&1, if {X}={Y} \\
&0, otherwise
\end{aligned}
\right. ,
\ 1_{Y}({X})=\left\{
\begin{aligned}
&1, if {X}\subseteq {Y} \\
&0, otherwise
\end{aligned}
\right.
\end{align}
\indent Suppose that the state vector $x$ in the space $\mathbb{X}$ is augmented with a unique label $\ell\in\mathbb{L}$, where $\mathbb{L}$ is a discrete label space, and $\textbf{X}$ represents the labeled RFS. 
Let $\mathcal{L}:\mathbb{L}\times\mathbb{X} \rightarrow \mathbb{L}$ be the projection $\mathcal{L}((x,\ell))=\ell$, $\mathcal{L}(\textbf{X})$ is the label set of $\textbf{X}$. The distinct label indicator $\Delta(\textbf{X})=\delta_{|\textbf{X}|}(\mathcal{L}(\textbf{X}))$ ensures the distinctness of the labels of $\textbf{X}$. All finite subsets of $\mathbb{L}$ are denoted by $\mathcal{F}(\mathbb{L})$. \\
\indent Augment the state with an unique label, the labeled multi-Bernoulli (LMB) RFS $\textbf{X}$ in the state space $\mathbb{X}$ and label space $\mathbb{L}$ can then be represented by the parameter set $\pi=\{{r^{(\ell)}, p^{(\ell)}}(x)\}_{\ell\in\mathbb{L}}$, and the density function is
\begin{equation}
\pi(\textbf{X})=\Delta(\textbf{X})\omega(\mathcal{L}(\textbf{X}))p^\textbf{X}
\end{equation}
where the weight
\begin{eqnarray}
\omega(L) &=& \prod\limits_{i\in\mathbb{L}}(1-r^{(i)})\prod\limits_{\ell\in L}\frac{1_{\mathbb{L}}(\ell)r^{(\ell)}}{1-r^{(\ell)}} 
\end{eqnarray} 

\indent Based on the labeled multi-Bernoulli RFS, an approximation of the multi-target Bayes filter was proposed in \cite{R19}, which consists of the following two steps: \\
\indent 1. Prediction: 
Suppose that the multi-target prior density and birth density are LMB RFSs. Then the predicted multi-target density is also a LMB RFS with state space $\mathbb{X}$ and label space $\mathbb{L}_+=\mathbb{L}\cup\mathbb{B}$ ($\mathbb{L}\cap\mathbb{B}=\varnothing$), where $\mathbb{L}$ and $\mathbb{B}$ are the label spaces of surviving and birth  target. This predicted density can be represented by the parameter set
\begin{equation}
\pi_+=\{(r_{+,S}^{(\ell)},p_{+,S}^{(\ell)})\}_{\ell\in\mathbb{L}}\cup\{(r_{B}^{(\ell)},p_{B}^{(\ell)})\}_{\ell\in\mathbb{B}}
\end{equation} 
where
\begin{eqnarray}
r_{+,S}^{(\ell)}&=&\eta_S(\ell)r^{(\ell)} \\
p_{+,S}^{(\ell)}&=&\langle p_s(\cdot,\ell)f(x|\cdot,\ell),p(\cdot,\ell) \rangle/\eta_S(\ell) \\
\eta_S(\ell)&=&\langle p_s(\cdot,\ell),p(\cdot,\ell) \rangle
\end{eqnarray}
Here, $p_s$ is the state dependent survival probability, and $f(x|\cdot,\ell)$ is the state transition density. $r_{B}^{(\ell)}$ and $p_{B}^{(\ell)}$ are the prior birth probability and state density of a new birth target, respectively. \\
\indent 2. Update: Suppose that the predicted multi-target LMB RFS is represented by the parameter set $\pi_+=\{(r_+^{(\ell)},p_+^{(\ell)})\}_{\ell\in\mathbb{L}_+}$ on $\mathbb{X}\times\mathbb{L}_+$. The multi-target predicted density can then be given by
\begin{equation}
\pi_+(\textbf{X}) = \Delta(\textbf{X})\sum\limits_{I\in \mathcal{L}(\textbf{X})}\omega_+^{(I_+)}\delta_{I_+}(\mathcal{L}(\textbf{X}))[p_+]^X
\end{equation}
where
\begin{equation}
\omega_+^{(I_+)}=\prod\limits_{\ell\in \mathbb{L}_+}(1-r_+^j)\prod\limits_{\ell\in I_+}\frac{1_{\mathbb{L}_+}(\ell)r_+^{(\ell)}}{1-r_+^{(\ell)}}
\end{equation}
After receiving the measurements, the LMB RFS that matches exactly the first moment of the multi-target posterior density can be denoted by the parameter set $\pi(\textbf{X}|Z)=\{(r^{(\ell)},p^{(\ell)}(x)\}_{\ell\in\mathbb{L}_+}$, in which, the updated existence probabilities $r^{(\ell)}$ and spatial distributions $p^{(\ell)}(x)$ of track $\ell$ are
\begin{eqnarray}
r^{(\ell)}&=&\sum\limits_{(I_+,\theta)\in \mathcal{F}(\mathbb{L_+})\times \Theta}\omega_+^{(I_+,\theta)}(Z)1_{I_+}(\ell) \\
p^{(\ell)}(x)&=&\frac{1}{r^{(\ell)}}\sum\limits_{(I_+,\theta)\in \mathcal{F}(\mathbb{L_+})\times \Theta}\omega_+^{(I_+,\theta)}(Z)1_{I_+}(\ell)p^{(\theta)}(x,\ell) \qquad
\end{eqnarray}
where $\Theta$ is the space of mappings $\theta:\mathbb{L}\rightarrow\{0,1,...,|Z|\}$, such that $\theta(i)=\theta(i')>0$ implies $i=i'$, and
\begin{eqnarray}
\omega^{(I_+,\theta)}(Z)&\propto&\omega_+(I_+)[\eta_Z^{(\theta)}]^{I_+} \\
p^{(\theta)}(x,\ell|Z)&=& \frac{p(x,\ell)\psi_Z(x,\ell;\theta)}{\eta_Z^{(\theta)}(\ell)} \\
\eta_Z^{(\theta)}(\ell)&=& \langle p(\cdot,\ell),\psi_Z(\cdot,\ell;\theta)\rangle \\
\psi_Z(x,\ell;\theta)&=& \delta_0(\theta(\ell))q_d(x,\ell) \\
&+&(1-\delta_0(\theta(\ell)))\frac{p_d(x,\ell)g(z_{\theta(\ell)}|x,\ell)}{\kappa(z_{\theta(\ell)})} \qquad
\end{eqnarray}
Here, $p_d(x,\ell)$ is the detection probability of the target, $q_d(x,\ell) = 1-p_d(x,\ell)$ is the probability for missed detection, $g(z_{\theta(\ell)}|x,\ell)$ is the measurement likelihood, and $\kappa(z_{\theta(\ell)})$ is the intensity of the clutter process. \\

\subsection{Conditional joint decision and estimation}
\indent The foundation of the CJDE method \cite{R10} is a novel Bayesian risk depends on the particular received measurement $z$. The decision and estimation costs are converted to a unified measurement by introducing additional weight coefficients $\{\alpha_{ij},\beta_{ij}\}$, that is
\begin{equation}
\bar{R}(z)\triangleq\displaystyle{\sum_i}\displaystyle{\sum_j}(\alpha_{ij}c_{ij}+\beta_{ij}E[C(x,\hat{x})|D^i,H^j,z])P\{D^i,H^j|z\}
\end{equation}
where $P\{D^i,H^j|z\}$ is the joint probability of decision and hypothesis, $c_{ij}$ is the cost of decision $D^i$ while the true hypothesis is $H^j$, and the conditional expected estimation cost $E[C(x,\hat{x})|D^i,H^j]=mse(\hat{x}|D^i,H^j)$ is the mean square error. The optimal solution is derived to minimize this new Bayes risk, the optimal decision $D$ is
\begin{equation}
D=D^i \qquad  if \qquad  C_C^i(z)\leq C_C^n(z) \quad \forall n
\end{equation}
where the posterior cost is
\begin{equation}
C_C^i(z)=\displaystyle{\sum_j}(\alpha_{ij}c_{ij}+\beta_{ij}E[C(x,\hat{x})|D^i,H^j,z])P\{H^j|z\}
\end{equation}
\indent To calculate $C_C^i(z)$ with $C(x,\hat{x})=\tilde{x}'\tilde{x}$, the key is to obtain the estimation cost $\epsilon_{ij}$. Assuming the optimal target estimate is 
\begin{equation}
\begin{aligned}
\hat{x}_n &= \sum\limits_{j}E(\hat{x}^{(j)}|H^{j},z)P\{H^{j}|z\}
\end{aligned}
\end{equation}
then and the estimation cost is
\begin{equation}
\begin{aligned}
\epsilon^{ij}(z)&\triangleq E[\tilde{x}'\tilde{x}|D^i,H^j,z] \\
&=mse(\hat{x}^{(ij)}|D^i,H^j,z)+E[(\hat{x}^{(ij)}-\hat{x})'(\cdot)|D^i,H^j,z] \\
&=mse(\hat{x}^{(j)}|H^j,z)+(\hat{x}^{(j)}-\check{x}^{(i)})'(\hat{x}^{(j)}-\check{x}^{(i)}), \forall z\in D^i
\end{aligned}
\end{equation}
\indent The recursive CJDE algorithm is shown as follows: \\
\indent (1) Initialize the parameters: $\hat{x}_{k-1}^{(j)}$, $P\{H^j|Z^{k-1}\}$ and so on. \\
\indent (2) Predict the state based on dynamics of $x_k$. Update $\hat{x}_k^{(j)}$ and $P\{H^j|Z^{k}\}$ by $z_k$. Then compute $\check{x}_k^{(i)}$ for decision $i$. \\
\indent (3) Compute $\epsilon^{ij}(Z^k)$ and get cost $C_C^i(Z^k)$. Then $D_k^i:C_C^i(Z^k)\leq C_C^n(Z^k),\forall n$. \\
\indent (4) Output the CJDE solution for time $k$. $D_k=D_k^i$ and $\hat{x}_k=\check{x}_k^{(i)}$. \\


\section{The recursive multi-target JDTC approach}
\vspace{-2pt}
\indent In this section, the mathematical formulation of the problem is firstly presented in 3.1. The multi-target JDTC algorithm and its Gaussian mixture (GM) implementation is then developed in 3.2 and 3.3, respectively. At last, the performance of the algorithm is analyzed in 3.4. 
\subsection{Problem formulation}
\indent Suppose that the class of a target is a time-invariant attribute, which can be distinguished according to the dynamic behavior. The target kinematic state at time $k$ for class $c_i$ can be modeled as
\begin{eqnarray}
x_{k} = F_{k-1,c}x_{k-1}+\Gamma_{k-1} w_{k-1,c}
\end{eqnarray}
where $F_{k-1,c}$ is the class-dependent state transition matrix, $w_{k,c}$ is the Gaussian process noise, and  $\Gamma_{k-1}$ is the gain matrix. The target can be observed by both Radar and ESM sensors, and the kinematic measurement of radar contains the range and angle measurements of the target, which can be given by 
\begin{eqnarray}
z_{k}^r& =& H_kx_k+v_{k}
\end{eqnarray}
where $H_k^r$ is the measurement matrix, and $v_k$ is the Gaussian noise with covariance $R_k$. The ESM sensors scan the frequency range to intercept emitted electromagnetic signals from the targets and identify the likely source emitters. The signal are processed and the angle of arrival can be obtained. The bearing measurement is
\begin{eqnarray}
z_{k}^e& =& H_k^ex_k+v_{k}^e
\end{eqnarray}
where $H_k^r$ and $v_k$ are the measurement matrix and Gaussian, respectively. Furthermore, the identification of the source emitters can be derived by sorting the received signals according to the radio frequency, signal parameters like modulation format, pulse repetition frequency, and so on. To account for the measurement error, the confusion matrix $\Pi$ can be defined. Assume there are $N$ types emitters, the matrix $\Pi$ contains $m\times m$ elements, where $m = 2^N$, and the element $\pi(i,j)$ in the matrix is the probability that
\begin{equation}
\pi_{ij} = \mathrm{Pr}\{\mathrm{declare} \quad E^j|\mathrm{ture} \quad E^i\} \qquad i,j=1,2,...,m 
\end{equation} 
\indent Assume that, at time $k$, $X_k=\{x_{k,1},...,x_{k,n}\}$ is the set of multi-target states, $Z_k^r=\{z_{k,1}^r,...,z_{k,m}^r,c_1,...,c_i\}$ is the set of noisy and cluttered measurements, where $\{z_{k,1},...,z_{k,m}\}$ is the measurement set generated from the targets and $\{c_1,...,c_i\}$ is the set of clutter. Similarly, $Z_k^e=\{z_{k,1}^e,...,z_{k,m}^e,c_1,...,c_i\}$ is measurement set of the ESM sensor, the measurement $z_{k,m}^e = [\beta_{k,m}, c_{k,m}]$ contains the angle of the target and the probability of the target type. The multi-target JDTC algorithm aims to estimate the target number and states, and determine their classes from a sequence of noisy and cluttered measurement sets. 

\subsection{The multi-target JDTC approach based on the generalized Bayesian risk}
\indent As multi-target JDTC is a dynamic problem and measurements are usually obtained sequentially, a new recursive Bayesian risk is firstly defined based on the labeled RFS. Suppose that $\mathcal{C}=\{\mathtt{C}_j\}_{j=1}^J$ is the class set which contains $J$ possible target classes, $\textbf{X}$ is the multi-target state RFS, $H^m = \{H_{\ell}^j\}_{\ell\in\mathcal{L}(\textbf{X})}$ and $D_k^n = \{D_{k,\ell}^i\}_{\ell\in\mathcal{L}(\textbf{X})}$ are the class hypothesis and decision sets of all the targets, respectively, where $H_\ell^j$ and $D_{k,\ell}^i$ are the class hypothesis and decision for track $\ell$. The new Bayesian risk is then given by
\begin{equation}
\begin{aligned}
\bar{R}_C(Z_k)&=\sum\limits_{m,n}\Big(\alpha_{mn}c_{mn}+\beta_{mn}E[C(\textbf{X},\hat{\textbf{X}})|D_k^{n},H^{m},Z_k] \\
&+\gamma_{mn}E[(|I_{mn}-\hat{I}|)|D_k^n,H^m,Z_k]\Big)P\{D_k^{n},H^{m}|Z_k\}
\end{aligned}
\end{equation}
where $c_{mn}$ is the cost of deciding on $D_k^n$ when the hypothesis $H^m$ is true, $C[(\textbf{X},\hat{\textbf{X}})|D_k^{n},H^{m},Z_k]$ is the conditional expected estimation cost of multi-target states, and $E[(|I_{mn}-\hat{I}|)|D_k^n,H^m,Z_k]$ is the conditional expected multi-target cardinality estimation error, $P\{D_k^{n},H^{m}|Z_k\}$ is the posterior probability of decision and hypothesis set, $\alpha_{mn}$, $\beta_{mn}$, and $\gamma_{mn}$ are the nonnegative weights used to unify the costs. \\
\indent To minimize $\bar{R}_C(Z_k)$, the optimal decision $D_k$ is
\begin{equation}
D_k=D_k^n \qquad  if \qquad  C_n(Z_k)\leq C_i(Z_k),\forall i
\end{equation}
where the cost $C_n(Z_k)$ for the decision $n$ is given by
\begin{equation}
\begin{aligned}
C_n(Z_k)=&\sum\limits_{m}\Big(\alpha_{mn}c_{mn}+\beta_{mn}E[C(\textbf{X},\hat{\textbf{X}})|D_k^n,H^{m},Z_k] \\
&+\gamma_{mn}E[(|I_{mn}-\hat{I}|)|D_k^n,H^m,Z_k]\Big)P_n(H^{m}|Z_k)
\end{aligned}
\end{equation}
Similar to (*), the decision conditioned estimation and costs are calculated using the measurements lie in the region of the decision region $\mathcal{D}^i_\ell$.
Based on the Bayes decision method, for target $\ell$, a set of Radar and ESM measurements $Z$ lie in the region of the decision region $\mathcal{D}^i_\ell$ when 
\begin{equation}
\quad C_i^{k}(Z|Z^{k-1})\leq C_n^{k}(Z|Z^{k-1}), \quad \forall n
\end{equation}
where, $C_i^{k}(Z|Z^{k-1})$ is the intermediate cost of target state estimation and classification
\begin{equation}
\begin{aligned}
&C_i^k(Z|Z^{k-1}) \\ 
&\qquad=\frac{1}{\rho}\sum\limits_{j}(\alpha_{ij}c_{ij}+\beta_{ij}\epsilon_{ij}^k)L(Z|Z^{k-1},H_\ell^j)P\{H_\ell^j|Z^{k-1}\}
\end{aligned}
\end{equation}
Here, $\rho$ is the normalization factor. $L(Z|Z^{k-1},H_\ell^j)$ is the likelihood functions conditioned on target type of $H_\ell^j$ of the kinematic and attribute measurements
\begin{equation}
\begin{aligned}
&L(Z|Z^{k-1},H_\ell^j) \\
&\qquad= f(z^r|Z^{k-1},H_\ell^l)f(z^e|Z^{k-1},H_\ell^j)\mathrm{Pr}(z_c=j|H_\ell^j)
\end{aligned}
\end{equation}
Especially, when The target belongs to two possible classes $H_\ell^j,j=1,2$ 
\begin{equation}
\frac{C_{1}}{C_{2}}\underset{D_1}{\overset{D_2}{\gtrless}} \frac{L(Z|Z^{k-1},H_\ell^j)P\{H_\ell^2|Z^{k-1}\}}{L(Z|Z^{k-1},H_\ell^j)P\{H_\ell^1|Z^{k-1}\}} 
\end{equation}
where $C_{1}=\alpha_{12} c_{12}+\alpha_{11} c_{11}+\beta_{11}\varepsilon_{11}+\beta_{12}\varepsilon_{12} $. 
Actually, the class decisions of each target form a partition of the measurement space. 
Here, the inclusion function $1_{\mathcal{D}_\ell^i}(z)$ is used in (*) to indicate whether the measurement $z$ lies inside the region $\mathcal{D}_\ell^i$. If $z\in \mathcal{D}_\ell^i$, $1_{\mathcal{D}_\ell^i}(z)=1$; if $z\notin \mathcal{D}_\ell^i$, $1_{\mathcal{D}_\ell^i}(z)=0$. When the measurement is missing, according to the mapping $\theta$, the corresponding likelihood function is equal to $1-p_d$. \\
\indent Assume at $k-1$, the posterior density of target $\ell$ can be given by
\begin{equation}
p_{k-1}(x,\ell) = \sum\limits_{j=1}^Jf_{k-1}(x,\ell|H_\ell^j)P(H_\ell^j)
\end{equation}
where $P(H_\ell^j)$ is the probability of the class hypothesis, and $f_{k-1}(x,\ell|H_\ell^j)$ is the class dependent target density. Then, the multi-target posterior density at $k-1$ can be represented as $\pi_{k-1}=\{(r_{k-1}^{(\ell)},p_{k-1}^{(\ell)}(\cdot|H_\ell^j),P(H_\ell^j))\}_{\ell\in\mathbb{L}}$. \\
\indent Suppose that the multi-target birth density is also LMB RFS with label set $\mathbb{B}$, the posterior density conditioned on the decision set $D_k^n = \{D_{k,\ell}^i\}_{\ell\in\mathcal{L}(\textbf{X})\cup\mathbb{B}}$ at time $k$ can be given by
\begin{equation}
\begin{aligned}
&\pi(\textbf{X}|D_k^n)= \\
&\frac{1}{\eta}\Delta(\textbf{X})\sum\limits_{I_{k-1},\theta}1_{\Theta(I_{k-1}\cup\mathbb{B})}(\theta)1_{\mathcal{D}_{k}^n}(Z_k^{(\theta)})\omega_{k-1}^{I_{k-1}}[\gamma_{Z_k}^{\theta}]^{I_{k-1}\cup\mathbb{B}} \\
&\qquad\times\left[\sum\limits_{j}p_n^{(\theta)}(\cdot,\ell|H_\ell^j,D_{k,\ell}^i,Z_k)P_n^{(\theta)}(H_\ell^j|D_{k,\ell}^i,Z_k)\right]^\textbf{X}
\end{aligned}
\end{equation}
where $\Theta$ is the space of mappings $\theta$ between the targets and the measurements from Radar and ESM sensors, i.e., $\theta:\mathbb{L}\rightarrow\{0,1,...,|Z_k^r|\}\times\{0,1,...,|Z_k^e|\}$, $\eta=\sum_{I_{k-1}}\sum_{\theta_k}1_{\mathcal{D}_{k}^n}(Z_k^{(\theta)})\omega_{k-1}^{I_{k-1}}[\gamma_{Z_k}^{\theta}]^{I_{k-1}\cup\mathbb{B}}$ is the normalization factor, and $1_{\mathcal{D}_{k}^n}(Z_k^{(\theta)})$ is the inclusion function that indicates whether the measurement lies inside the region of corresponding decision $D_{k,\ell}^i$ according to the mapping. The posterior density and class probability of each target can be calculated as
\begin{align}
p_n^{(\theta)}(x,\ell|H_\ell^j,D_{k,\ell}^i,Z_k) &= \frac{1_{\mathcal{D}_\ell^i}(z_{\theta(\ell)})\psi_Z(x,\ell;\theta)f_{k|k-1}(x,\ell)p_{k-1}^{(\ell)}(x|H_\ell^j)}{\eta_Z^{(\theta)}(\ell|D_{k,\ell}^i,H_{\ell}^j)} \quad \\
P_n^{(\theta)}(H_\ell^j|D_{k,\ell}^i,Z_k) &= \frac{ \eta_Z^{(\theta)}(\ell|D_{k,\ell}^i,H_\ell^j)P_{k-1}(H_\ell^j)}{\sum_{j}\eta_Z^{(\theta)}(\ell|D_{k,\ell}^i,H_\ell^j)P_{k-1}(H_\ell^j)} \\
\eta_Z^{(\theta)}(\ell|D_{k,\ell}^i,H_\ell^j) &= \langle \Psi_Z(x,\ell;\theta),f_{k|k-1}(x,\ell)p_{k-1}^{(\ell)}(x|H_\ell^j) \rangle \\
\Psi_Z(x,\ell;\theta) &= \psi_z^r(x,\ell;\theta)\psi_z^e(x,\ell;\theta) 
\end{align}
where $f_{k|k-1}(x,\ell)$ is the state transition function, $\psi_z^r(x,\ell;\theta)$, $\psi_z^e(x,\ell;\theta)$ are the likelihood functions of the Radar and ESM measurements, respectively.
\begin{eqnarray}
\psi_z^r(x,\ell;\theta) &=& \left\{
\begin{aligned}
&1-p_d^r(x,\ell), \qquad \qquad z_{\theta(\ell)}=\varnothing \\
&\frac{p_d^r(x,\ell)g(z_{\theta(\ell)}|x,\ell)}{\kappa(z_{\theta(\ell)})}, other
\end{aligned}
\right. \\
\psi_z^e(x,\ell;\theta) &=& \left\{
\begin{aligned}
&1-p_d^e(x,\ell), \qquad \qquad z_{\theta(\ell)}=\varnothing \\
&\frac{p_d^e(x,\ell)g(z_{\theta(\ell)}|x,\ell)\mathrm{Pr}(z_c=j|H_\ell^j)}{\kappa(z_{\theta(\ell)})}, other
\end{aligned}
\right. 
\end{eqnarray}
\indent In (*), the weights $\omega_{k}^{I_{k}}$ is equal to $1_{\mathcal{D}_{k}}(Z_k^\theta)\omega_{k-1}^{I_{k-1}}[\gamma_{Z_k}^{\theta}]^{I_{k-1}\cup\mathbb{B}}$, where
\begin{eqnarray}
\gamma_{Z_k}^\theta(\ell) = \left\{
\begin{aligned}
1-p_s(x,\ell),\qquad \forall \ell\in I_{k-1}, \ell \notin I_k \\
p_s(x,\ell)\eta_{z_k}^\theta(\ell),\qquad \forall \ell\in I_{k-1}, \ell \in I_k \\
1-r(\ell),\qquad \forall \ell\in \mathbb{B}, \ell \notin I_k \\
r(\ell)\eta_{z_k}^\theta(\ell),\qquad \forall \ell\in \mathbb{B}, \ell \in I_k \\
\end{aligned}
\right.
\end{eqnarray}

The LMB RFS that matches exactly the first moment of the multi-target posterior density can then be given by
\begin{equation}
\pi_k^n(x|Z_k)=\{(r_n^{(\ell)},p_n^{(\ell)}(x|H_\ell^j)P_n(H_\ell^j))\}_{\ell\in\mathbb{L}_+}
\end{equation}
where
\begin{align}
r_n^{(\ell)}&=\sum\limits_{I_{k},\theta}\omega_n^{(I_{k},\theta)}(Z_k)1_{I_{k}}(\ell) \\
p_n^{(\ell)}(x|H_\ell^j)&=\frac{1}{r^{(\ell)}}\sum\limits_{I_{k},\theta}\omega_n^{(I_{k},\theta)}(Z_k)1_{I_{k}}(\ell)p_n^{(\theta)}(x,\ell|H_\ell^j) \\
P_n(H_\ell^j) &= \frac{1}{r^{(\ell)}}\sum\limits_{I_{k},\theta}\omega_n^{(I_{k},\theta)}(Z_k)1_{I_{k}}(\ell)P_n^{(\theta)}(H_\ell^j) 
\end{align}
\indent In the update step, the multi-target posterior density is computed conditioned on the decision. Additionally, multi-target distribution is approximated by preserving the spatial density of each track with exact match of the first moment. \\
\indent To derive the optimal CJDE solution, the costs of multi-target detection, tracking and classification need to be calculated. For the exact calculation of the posterior density for each target involving the MTA's, the CJDE cost can be calculated as
\begin{equation}
\begin{aligned}
&C_n(Z_k) \\
&\quad= \sum\limits_{m}\left(\sum\limits_{c\in\mathbb{C}}\omega_n^c(\alpha_{mn}c_{mn}+\beta_{mn}\varepsilon_{X})+\gamma_{mn}\varepsilon_{I}\right)P_n(H^{m}|Z_k) \\
\end{aligned}
\end{equation}
where $c\in\mathbb{C}$ represents $(I_{k},\theta)\in\mathcal{F}(\mathbb{L})\times\Theta$, 
and the hypothesis probability
\begin{equation}
P_n(H^{m}|Z_k)=\prod\limits_{\ell\in I_{k}}P_n(H_\ell^j)
\end{equation} 
The calculation of the CJDE cost can be divided into two parts. Firstly, the joint cost of target detection and classification can be calculated as
\begin{equation}
\begin{aligned}
&\tilde{C}_n(Z_k)= \\
&=\sum\limits_{m}\sum\limits_{c}\omega_n^c(\alpha_{mn}c_{mn}+\beta_{mn}\varepsilon_{X})P_n(H^{m}|Z_k) \\
&=\sum\limits_{c}\sum\limits_{m}\sum\limits_{\ell\in I_{k}}\omega_n^{c}(\alpha_{mn}^\ell c_{k,\ell}^{ij}+\beta_{mn}^\ell \varepsilon_{k,\ell})\prod\limits_{\ell\in I_{k}}P_n(H_\ell^j) \\
&=\sum\limits_{c}\sum\limits_{m}\sum\limits_{\ell\in I_{k}}\omega_n^{c}(\alpha_{mn}^\ell c_{k,\ell}^{ij}+\beta_{mn}^\ell \varepsilon_{k,\ell})P_n(H_\ell^j)\prod\limits_{I_{k}\backslash\ell}P_n(H_\ell^j) \\
&=\sum\limits_{c}\sum\limits_{m}\sum\limits_{\ell\in I_{k}}\omega_n^{c}(\alpha_{mn}^\ell c_{k,\ell}^{ij}+\beta_{mn}^\ell \varepsilon_{k,\ell})P_n(H_\ell^j)
\end{aligned}
\end{equation}
where $c_{k,\ell}^{ij}$ is the cost of deciding on $D_{k,\ell}^i$ when hypothesis $H_\ell^j$ is true for track $\ell$, and the term $\varepsilon_x$ denotes the estimation cost of target state, which can be calculated as (*) 
\begin{equation}
\begin{aligned}
\varepsilon_x &=E[C(x_\ell,\hat{x}_\ell)|D_{k,\ell}^i,H_\ell^{j},Z_k] \\
&=mse(\hat{x}_{k,\ell}^{ij})+(\hat{x}_{k,\ell}^{ij}-\check{x}_{k,\ell}^{i})^\mathrm{T}(\hat{x}_{k,\ell}^{ij}-\check{x}_{k,\ell}^{i})
\end{aligned}
\end{equation}
where $\hat{x}_{k,\ell}^{ij}$ is the class dependent state estimate derived with respect to the posterior distribution, and $\check{x}_{k,\ell}^{i}$ is the optimal estimate for the decision $D_{k,\ell}^i$, which can be calculated as
\begin{equation}
\check{x}_{k,\ell}^{i} = \sum\limits_{j=1}^J\hat{x}_{k,\ell}^{ij}P_n(H_\ell^j), \quad z_{\theta(\ell)}\in \mathcal{D}_{k,\ell}^i
\end{equation}
If no measurements lie inside the region of the decision $D_{k,\ell}^i$, the estimation cost can be computed by replacing the estimate $\hat{x}_{k,\ell}^{ij}$ with the prediction. \\
\indent 
Because the original LMB filter propagates multi-target density with an exact match of the first posterior moment, it does not exhibit a cardinality bias \cite{R19}. Therefore, the multi-target cardinality estimate of the original LMB filter is used as the optimal estimate here. As given in (36)-(41), the posterior multi-target cardinality estimates is dependent on the decision because the calculation of the weight involving the inclusion function $1_{\mathcal{D}_\ell^i}(z)$. Therefore, the coefficients $\gamma_{ij}$ of the existence probability estimation costs are reasonable set to be equal for all class hypotheses, i.e., $\gamma_{mn}=\gamma_m$ for all $H^m$. In this case, the multi-target cardinality estimation cost can be calculated as
\begin{equation}
\begin{aligned}
\varepsilon_I 
&=\sum\limits_{N}\sum\limits_{(I,\theta)\in\mathcal{F}_N(\mathbb{L})\times\Theta}N\left(\omega^{(I,\theta)}-\omega_{ij}^{(I,\theta)}\right)
\end{aligned}
\end{equation}
Calculate the cost using (45)-(48), then, the optimal decision is $D_k^n:C_n(Z_k)\leq C_i(Z_k), \forall i$, and the corresponding target state estimates are derived using the conditional LMB filter. \\
\indent The proposed recursive multi-target JDTC algorithm is summarized as follows:  
\begin{algorithm}{The Recursive CJDE-LMB Algorithm \\}
\indent 1. Predict prior multi-target density using the class-dependent dynamic model according to the hypothesis. \\
\indent 2. Update $\check{x}_{k,\ell}^i$, $P_n(H_\ell^j)$ and $\omega_n^{(I_+,\theta)}$ for decision $D_k^n$ using the conditional LMB filter. \\ 
\indent 3. Calculate the joint detection, tracking and classification cost $C_n(Z^k)$ using (45)-(48), and the optimal decision is then $D_k^n:C_n(Z_k)\leq C_i(Z_k), \forall i$. \\
\indent 4. Output the CJDE solution for time k: the optimal decision $D_k=D_k^n$, the target existence probability $r_n^{(\ell)}$ and the state estimate $\check{x}_{k,\ell}^{i}$. \\
\end{algorithm} 

\subsection{Gaussian mixture implementation}
\indent In this subsection, the Gaussian mixture implementation of the proposed recursive JDTC approach is developed. \\
\indent 1) Prediction:
Suppose that at time $k-1$, the multi-target density can be represented as $\pi_{k-1}(\textbf{X})=\{(r_{k-1}^{(\ell)},p_{k-1}^{(\ell)}(x|H_\ell^j)P(H_\ell^j))\}_{\ell\in\mathbb{L}}$, where $p_{k-1}^{(\ell)}(x|H_\ell^j)$ is the density of track $\ell$ that can be typically modeled by a Gaussian mixture
\begin{equation}
p_{k-1}^{(\ell)}(x|H_\ell^j)=\sum\limits_{n=1}^{N_{k-1,\ell}^j}\omega_{k-1,j}^{(n)}\mathcal{N}(x,m_{k-1,j}^{(n)},P_{k-1,j}^{(n)})
\end{equation}
where $m_{k-1,j}^{(n)}$ and $P_{k-1,j}^{(n)}$ are the mean value and covariance of the state vector, the predicted multi-target density can then be represented as (29).
Suppose that the predicted multi-target density can be represented by the parameters $\pi_{k|k-1}(\textbf{X})=\{(r_{k|k-1}^{(\ell)},p_{k|k-1}^{(\ell)}(x|H_\ell^j)P(H_\ell^j))\}_{\ell\in\mathbb{L}_+}$, where the density $p_{k|k-1}^{(\ell)}(x|H_\ell^j)$ can be represented by a Gaussian mixture as 
\begin{equation}
p_{k|k-1}^{(\ell)}(x|H_\ell^j) = \sum\limits_{n=1}^{N_{k|k-1,\ell}^j}\mathcal{N}(x;{m}_{k|k-1,j}^{(n)},{P}_{k|k-1,j}^{(n)})
\end{equation}
When the measurement set $Z_k$ is collected at time $k$, the posterior multi-target density conditioned on the decision $\{D_k^n\}$ is 
\begin{equation}
\begin{aligned}
\pi_k^n(\textbf{X}|Z_k)=\Delta(\textbf{X})\sum\limits_{(I_+,\theta)\in\mathcal{F}(\mathbb{L}_+)\times\Theta}\omega_n^{(I_+,\theta)}(Z_k)\delta_{I_+}(\mathcal{L}(\textbf{X}))&\\
\times\left[\sum\limits_{j}p_n^{(\theta)}(\cdot,\ell|H_\ell^j,D_{k,\ell}^i,Z_k)P_n^{(\theta)}(H_\ell^j|D_{k,\ell}^i,Z_k)\right]^\textbf{X}&
\end{aligned}
\end{equation}
where the weight 
\begin{align}
\omega_n^{(I_+,\theta)}(Z_k)&\propto\omega_+^{(I_+)}\left[\eta_Z^{(\theta)}(\ell|D_{k,\ell}^i,H_\ell^j)\right]^{I_+}
\end{align}
and
\begin{align}
\nonumber&\eta_Z^{(\theta)}(\ell|D_{k,\ell}^i,H_\ell^j) \\
\nonumber&= 1_{\mathcal{D}_\ell^i}(z_{\theta(\ell)})\Big((1-p_d)+p_d\frac{1}{\lambda c(k)} \\ &\times\sum\limits_{n=1}^{N_{k|k-1,\ell}^j}\omega_{k|k-1,j}^{(n)}\mathcal{N}(z;H_km_{k|k-1,j}^{(n)},H_kP_{k|k-1,j}^{(n)}H_k^\mathrm{T}+R_k)\Big)
\end{align}
The posterior density of each target can be calculated using the measurement augmented optimal Kalman filtering method as follows
\begin{align}
\nonumber p_n^{(\theta)}(x,\ell|Z_k) &=  \sum\limits_{n=1}^{N_{k|k-1,\ell}^j}\omega_{k|k-1}^{(n)}\big((1-p_d)\mathcal{N}(x;m_{k|k-1,j}^{(n)},P_{k|k-1,j}^{(n)}) \\
&\qquad \qquad \qquad+ p_dq_{k,j}^{(n)}(z_{\theta(\ell)})\mathcal{N}(x;{m}_{k,j}^{(n)},{P}_{k,j}^{(n)})
\big)
\end{align}
where
\begin{align}
m_{k|k-1,j}^{(n)} &= F_{k|k-1}^jm_{k-1,j}^{(n)} \\
P_{k|k-1,j}^{(n)} &= F_{k|k-1}^jP_{k-1,j}^{(n)}(F_{k|k-1}^j)^\mathrm{T}+Q_{k|k-1}^j \\
q_{k,j}^{(n)}(z_{\theta(\ell)}) &= \mathcal{N}(z_{\theta(\ell)};H_km_{k|k-1,j}^{(n)},P_{k|k-1,j}^{(n)}) \\
m_{k,\ell}^{(n)} &= \hat{x}_{k|k-1,\ell}+\textbf{K}_{k}(\textbf{z}_{\vartheta(\ell)}-\textbf{z}_{+}) \\
\textbf{z}_{+} &= \textbf{H}\hat{x}_{k|k-1,\ell}+\textbf{b} \\
\textbf{K}_{k,j}^{(n)} &= P_{k|k-1}\textbf{H}^\mathrm{T}[\textbf{H}P_{k|k-1,\ell}\textbf{H}^\mathrm{T}+\textbf{R}]^{-1} \\
P_{k,j}^{(n)} &= (I-\textbf{K}_{k}\textbf{H})P_{k|k-1} \\
\textbf{S}_{k} &= \textbf{H}P_{k|k-1}\textbf{H}^\mathrm{T}+\textbf{R}
\end{align}
where $\textbf{z}_{\vartheta(\ell)}=[z_{\theta_k^1(\ell)}^\mathrm{T},...,z_{\theta_k^s(\ell)}^\mathrm{T}]^\mathrm{T}$ represents the augmented measurements of $s$ sensors, and $\textbf{H}=[H_1^\mathrm{T},...,H_n^\mathrm{T}]^\mathrm{T}$ and $\textbf{R}=\mathrm{diag}(R_1,...,R_n)$ are corresponding augmented measurement and covariance matrices. Then the approximated target density can be derived using (41)-(43). \\

3) Calculate the risk: Compute the class dependent posterior estimate and associated covariance with respect to the distribution given in (60), that is
\begin{eqnarray}
\hat{x}_{k,\ell}^{ij} &=& \sum\limits_{n=1}^{N_{k,\ell}^{j}}\omega_{k,ij}^{(n)}m_{k,ij}^{(n)} \\
P_{k,\ell}^{ij} &=& \sum\limits_{n=1}^{N_{k,\ell}^{j}}\omega_{k,ij}^{(n)}\left(P_{k,ij}^{(n)}+(m_{k,ij}^{(n)}-\hat{x}_{k}^{ij})(m_{k,ij}^{(n)}-\hat{x}_{k}^{ij})^\mathrm{T}\right)
\end{eqnarray}
Then, the optimal estimate of track $\ell$ is
\begin{equation}
\begin{aligned}
\check{x}_{k,\ell}^i = \sum\limits_{j=1}^{J}\hat{x}_{k,\ell}^{ij}P_k^i(H_\ell^j) \\
\end{aligned}
\end{equation}
For the explicit Gaussian mixture implementation of the conditioned LMB filter, the estimation cost $\varepsilon_{X}$ in (45) can be given by
\begin{equation}
\begin{aligned}
\varepsilon_{X}&=\sum\limits_{\ell\in\mathcal{L}(\textbf{X})}\left(tr(P_{k,\ell}^{ij})+(\hat{x}_{k,\ell}^{ij}-\check{x}_{k,\ell}^i)^\mathrm{T}(\hat{x}_{k,\ell}^{ij}-\check{x}_{k,\ell}^i)\right)
\end{aligned}
\end{equation}
Finally, compute the CJDE cost for decision $D_k^n$ using (45)-(49), then the optimal solution can be derived.


\subsection{Performance analysis}
Because the detection of the target is the prerequisite of tracking and classification, if $\gamma_i$ is relative small, the CJDE cost $C_{m}(Z)\approx\sum_{\ell\in I_k}\alpha_{mn}^\ell c_{mn}^\ell+\beta_{mn}^\ell\varepsilon_{x}^\ell$. Because the estimation and classification costs in the Bayes risk are nonnegative, in this case, the target tends to be judged as missed for less state estimation and classification costs, and an incorrect JDTC solution maybe derived. 
Assume that no measurements lie inside the region of $D_\ell^i$, the weight is nonnegative the existence probability of the target is 
\begin{align}
r_n^{(\ell)}&=\sum\limits_{I_{k},\theta}\omega_n^{(I_+,\theta)}(Z_k)1_{I_+}(\ell) \\
&=\sum\limits_{I_{k-1}\cup\mathbb{B},\theta}1_{I_+}(\ell)\frac{1_{\mathcal{D}_{k}}(Z_k^\theta)\omega_{k-1}^{I_{k-1}}[\gamma_{Z_k}^{\theta_k}]^{I_{k}\cup\mathbb{B}}}{\sum\limits_{I_k,\theta_k}1_{\mathcal{D}_{k}}(Z_k^\theta)\omega_{k-1}^{I_{k-1}}[\gamma_{Z_k}^{\theta_k}]^{I_{k-1}\cup\mathbb{B}}} \\
&=(1-r_{k-1})r(\ell)\eta_{z_k}^\theta(\ell) \\
&+r_{k-1}\frac{p_s(x,\ell)p_d^r(x,\ell)p_d^e(x,\ell)}{p_s(x,\ell)(1-p_d^r(x,\ell)p_d^e(x,\ell))+(1-p_s(x,\ell))} 
\end{align}
Therefore, $\gamma_i$ can be chosen to make the maximum cost of the target detection approximate equal to the sum of the maximum costs of estimation and classification, i.e., $\gamma_i\approx(\alpha_{mn}\cdot1+\beta_{mn}\cdot max(\varepsilon_{x}))/(1-\bar{p})$, where $\bar{p}$ is the target existence probability estimate calculated with an empty set of measurements. In this case, the target detection cost will be predominant and the multi-target JDTC problem is solved with optimal estimate of the target number. 

\section{Simulations}
\indent In this section, numerical examples are presented to illustrate the effectiveness and superiority of the proposed CJDE-LMB algorithm. In addition, the results derived with different parameters are also compared.
\subsection{Example 1}
\indent Suppose that there are several targets with two possible classes move in a two-dimensional scenario. The classes differ from each other in terms of the dynamic behaviors, each class has a corresponding set of possible motion models. The $i$th model for class $j$ is
\begin{equation}
x_k = F_{k,i}x_{k-1}+w_{k,i}
\end{equation}
where $F_{k,i}$ is the model-dependent state transition matrix, and $w_{k,i}$ is Gaussian noise with covariance $Q_{k,i}$. The target of class 1 only has the constant velocity (CV) model with the following parameters 
\begin{equation}
\begin{aligned}
F_{k,1}&=diag\left( \begin{bmatrix} 1&T \\ 0&1 \end{bmatrix}, \begin{bmatrix} 1&T \\ 0&1 \end{bmatrix} \right) \\
Q_{k,1}&=diag\left( \begin{bmatrix} T^2&T \\ T&1 \end{bmatrix}, \begin{bmatrix} T^2&T \\ T&1 \end{bmatrix} \right)\sigma_v^2
\end{aligned}
\end{equation}
where $\sigma_v$ is the process noise with the covariance $\sigma_v^2=1\ \mathrm{m^2/s^2}$. \\
\indent The target of class 2 has two possible dynamic models, the CV model as before, and the constant accelerate (CA) model with parameters
\begin{equation}
\begin{aligned}
F_{k,2}&=diag\left( \begin{bmatrix} 1&T&\frac{1}{2}T^2 \\ 0&1&T \\ 0&0&1 \end{bmatrix}, \begin{bmatrix} 1&T&\frac{1}{2}T^2 \\ 0&1&T \\ 0&0&1 \end{bmatrix} \right) \\
Q_{k,2}&=diag\left( \begin{bmatrix} \frac{1}{4}T^4&\frac{1}{2}T^3&\frac{1}{2}T^2 \\ \frac{1}{2}T^3&T^2&T \\ \frac{1}{2}T^2&T&1 \end{bmatrix}, \begin{bmatrix} \frac{1}{4}T^4&\frac{1}{2}T^3&\frac{1}{2}T^2 \\ \frac{1}{2}T^3&T^2&T \\ \frac{1}{2}T^2&T&1 \end{bmatrix} \right)\sigma_a^2
\end{aligned}
\end{equation}
where $\sigma_a$ is the process noise with the covariance $\sigma_a^2=10\ \mathrm{m^2/s^4}$. The model transition probability matrix is set as
\begin{equation}
\pi=\begin{bmatrix} 0.7\quad&0.3 \\ 0.3\quad&0.7 \end{bmatrix}
\end{equation}
\indent The kinematic measurement is $z_k=[x_k,y_k]^\mathrm{T}+w_k$, where $[x_k,y_k]$ is the position of the target, and $w_k$ is the Gaussian measurement noise with the covariance $R_k=diag[\sigma_x^2,\sigma_y^2]$, $\sigma_x = \sigma_y = 2\ \mathrm{m}$. The target detection probability $p_d=0.98$, and the intensity of the Poisson distributed clutter is $6\times10^{-5}$. \\
\indent In the scenario, there are two non-maneuvering targets and one maneuvering target move within the two-dimensional scenario. Target 1 moves straight from the beginning to the end, with the initial location $[-200,700]\ \mathrm{m}$ and velocity $[50,0]\ \mathrm{m/s}$. Target 2 appears at $k=5$ with the initial location $[-200,1000]\ \mathrm{m}$, and moves straight with constant velocity $[40,30]\ \mathrm{m/s}$ until it disappears at $k=25$. The maneuvering target 3 appears at the $k=3$ and disappears at $k=27$. It moves straight from location $[0,1900]\ \mathrm{m}$ with a constant acceleration of $[4,-3]\ \mathrm{m/s}^2$. \\
\indent The multi-target detection, tracking, and classification performance of the CJDE-LMB algorithm is compared with the traditional methods in terms of the multi-target cardinality estimates, optimal subpattern assignment (OSPA) distance \cite{R20}, and the probability of correct classification, respectively. Moreover, the overall performance is evaluated by the joint performance metric (JPM), which is calculated with the costs of target detection, tracking, and classification. \\
\indent The compared methods are the follows: \\
\indent 1) Estimation-Then-Decision: The target state is first estimated using the GNN approach, and the decision is then made based on the ratio of current measurement likelihoods of the predicted states conditioned on different hypotheses. \\
\indent 2) Decision-Then-Estimation: The target class is first determined, which minimizes the Bayes decision risk, and the target state is then estimated given the decided class. \\
\indent 3) Estimate the joint target state-class probability density: As proposed in \cite{R01}, the class-dependent posterior density is firstly calculated using the particle implementation of the PHD filter with corresponding dynamic models. Then, the target state and class probabilities are obtained by clustering the particles. This method is referred to as YW-JDTC here. \\
\indent In the simulation, the target survival probability is $p_s=0.98$, and the target birth probability is $p_b=0.02$. The density of the new birth target is $b_k=\mathcal{N}(x;m_b,Q_b)$, where the parameters $m_{\gamma,k}^1 = [-200,50,0,700,0,0]^\mathrm{T}$, $m_{\gamma,k}^2 = [-200,40,0,1000,30,0]^\mathrm{T}$, and $m_{\gamma,k}^3 = [0,20,4,1900,-15,-3]^\mathrm{T}$, while the state covariances are $P_{\gamma,k}^1 = P_{\gamma,k}^2 = P_{\gamma,k}^3 = diag([100, 10, 1, 100, 10, 1])$. All the classes have an equal initial probability, and the initial probabilities of the two models for the maneuvering hypothesis are equal to 0.5. 
According to the guidance of parameter choice provided before, the parameters in the new CJDE risk are set to be $\alpha_{mn}^1=20, \beta_{mn}^1=1, \gamma_{mn}^1=100$. 
The simulation results are obtained over 1000 Monte Carlo trials. \\

\begin{figure*}[!htp]
	\centering
	\subfigure[Cardinality estimate]{
		\label{fig_aerial_org}
		\includegraphics[width=2.5in]{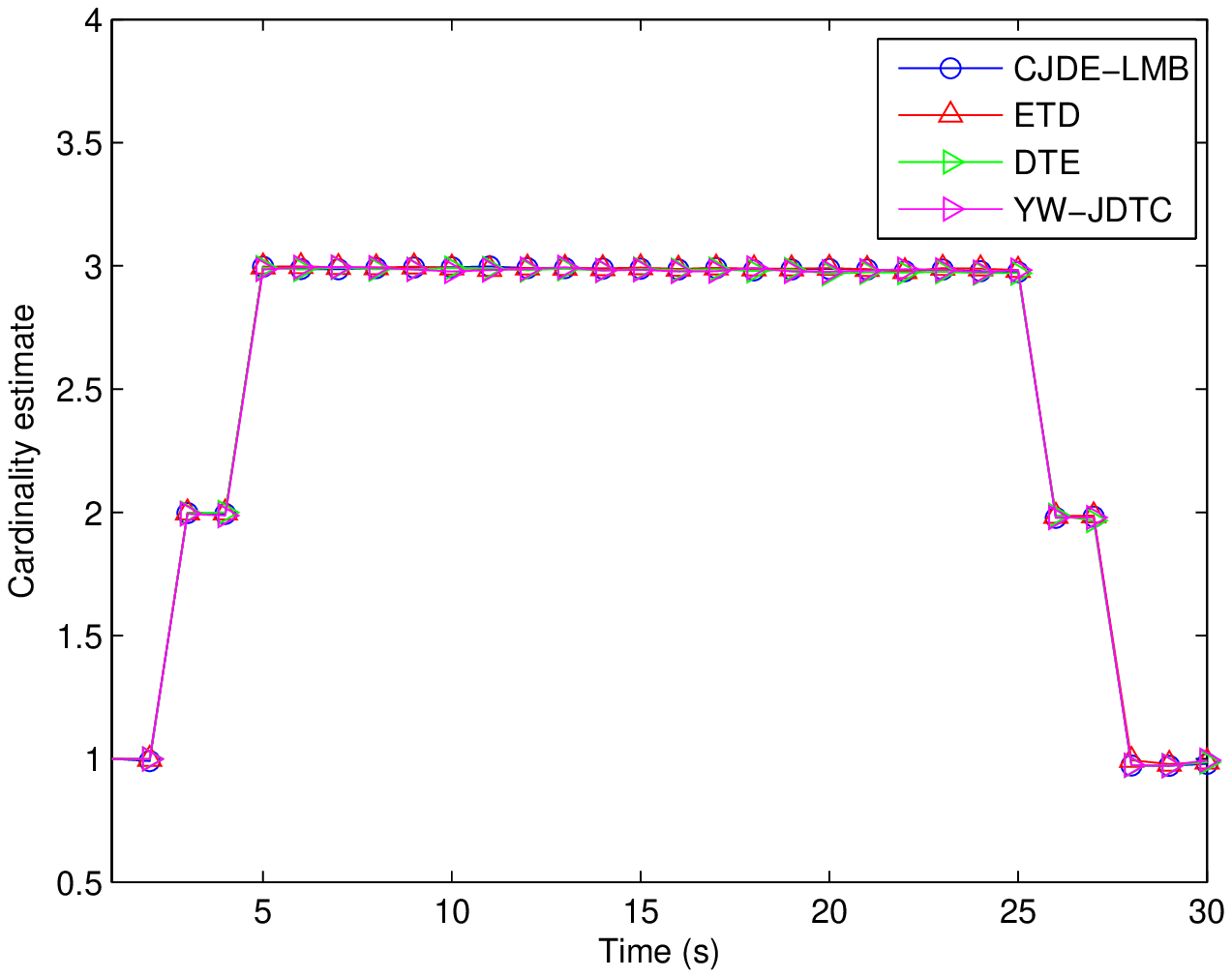}
	}
	\subfigure[OSPA distance]{
		\label{fig_barbara_org}
		\includegraphics[width=2.5in]{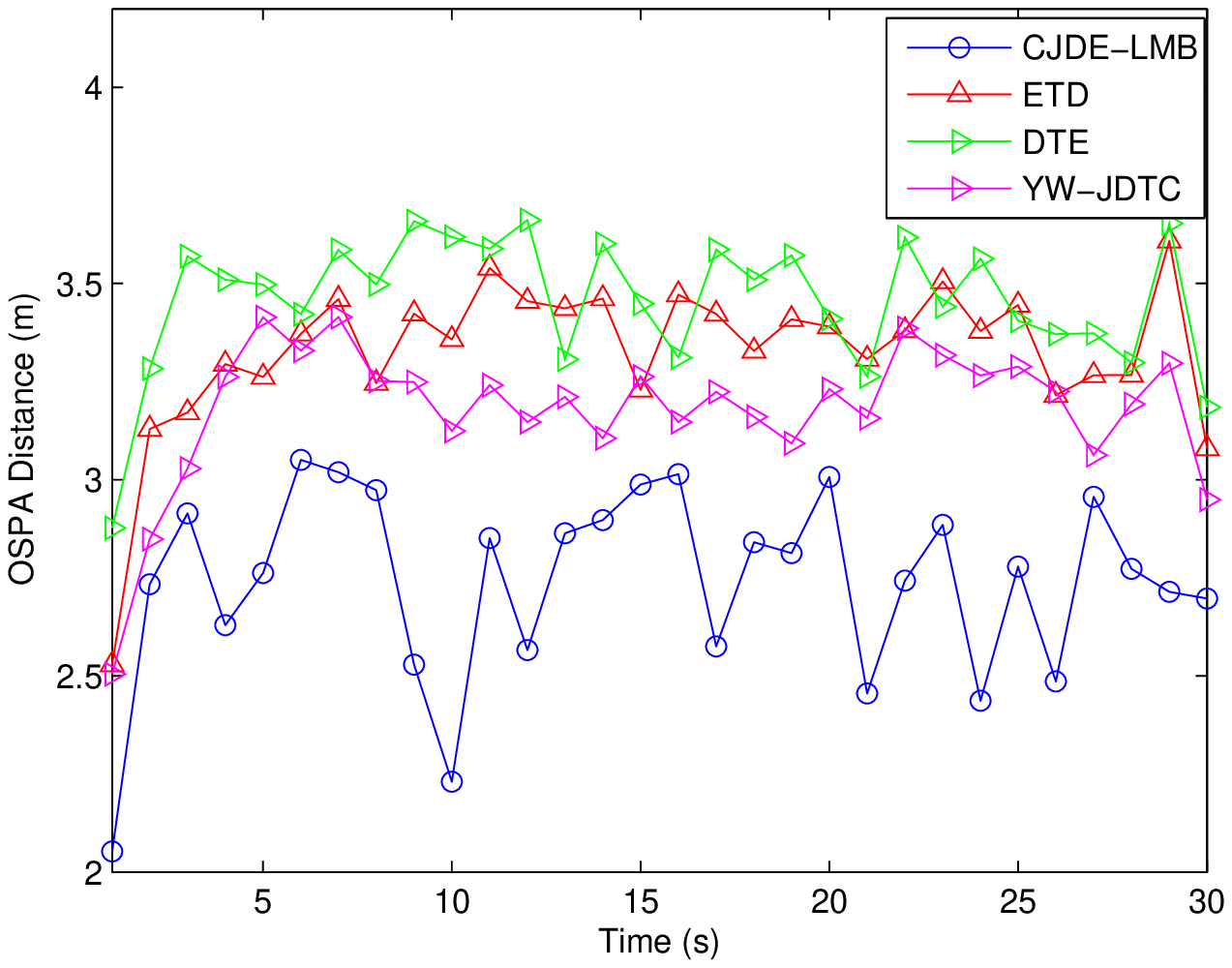}
	}
	\subfigure[Average probability of incorrect Classification]{
		\label{fig_boat_org}
		\includegraphics[width=2.5in]{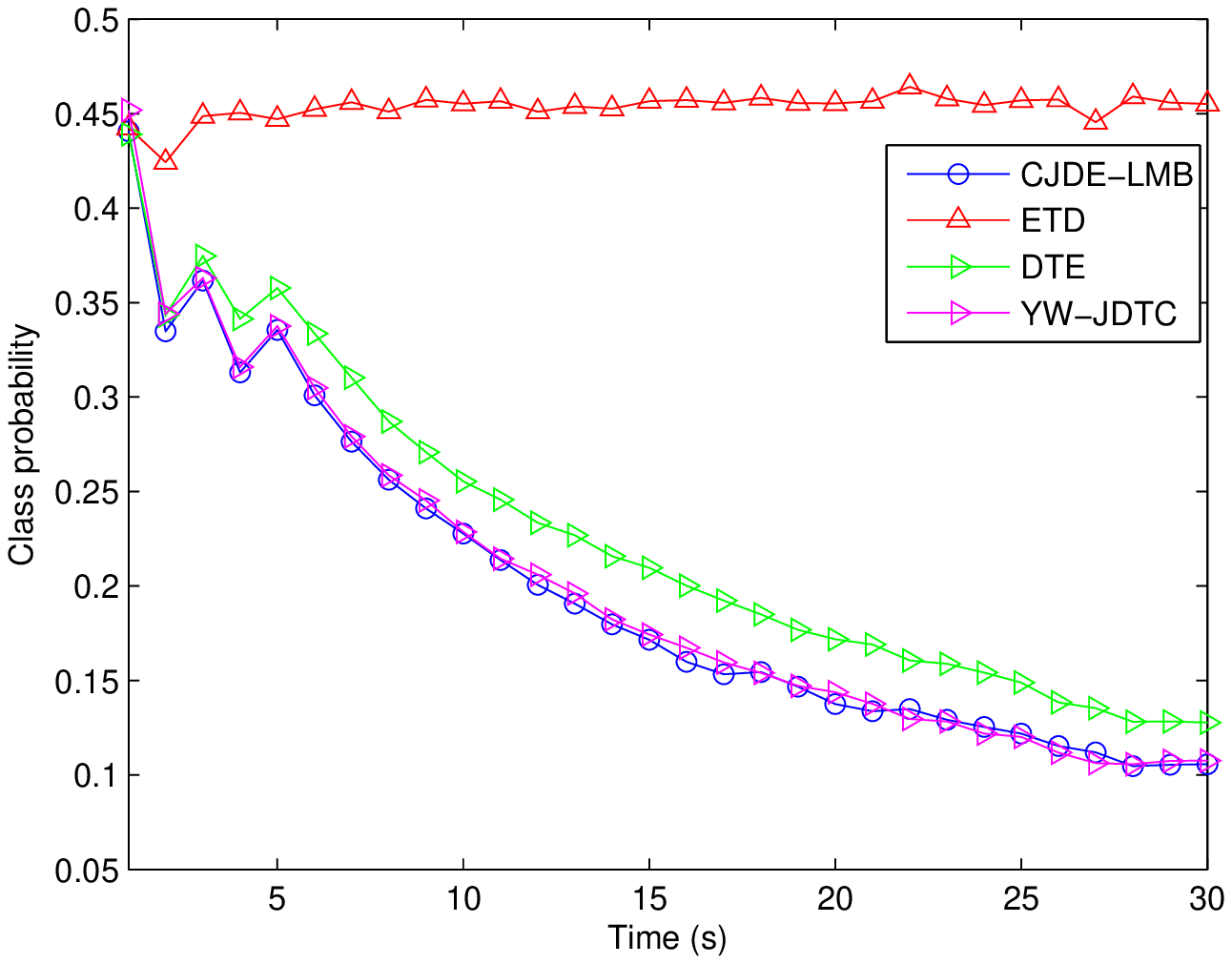}
	}
	\subfigure[Joint performance metric]{
		\label{fig_couple_org}
		\includegraphics[width=2.5in]{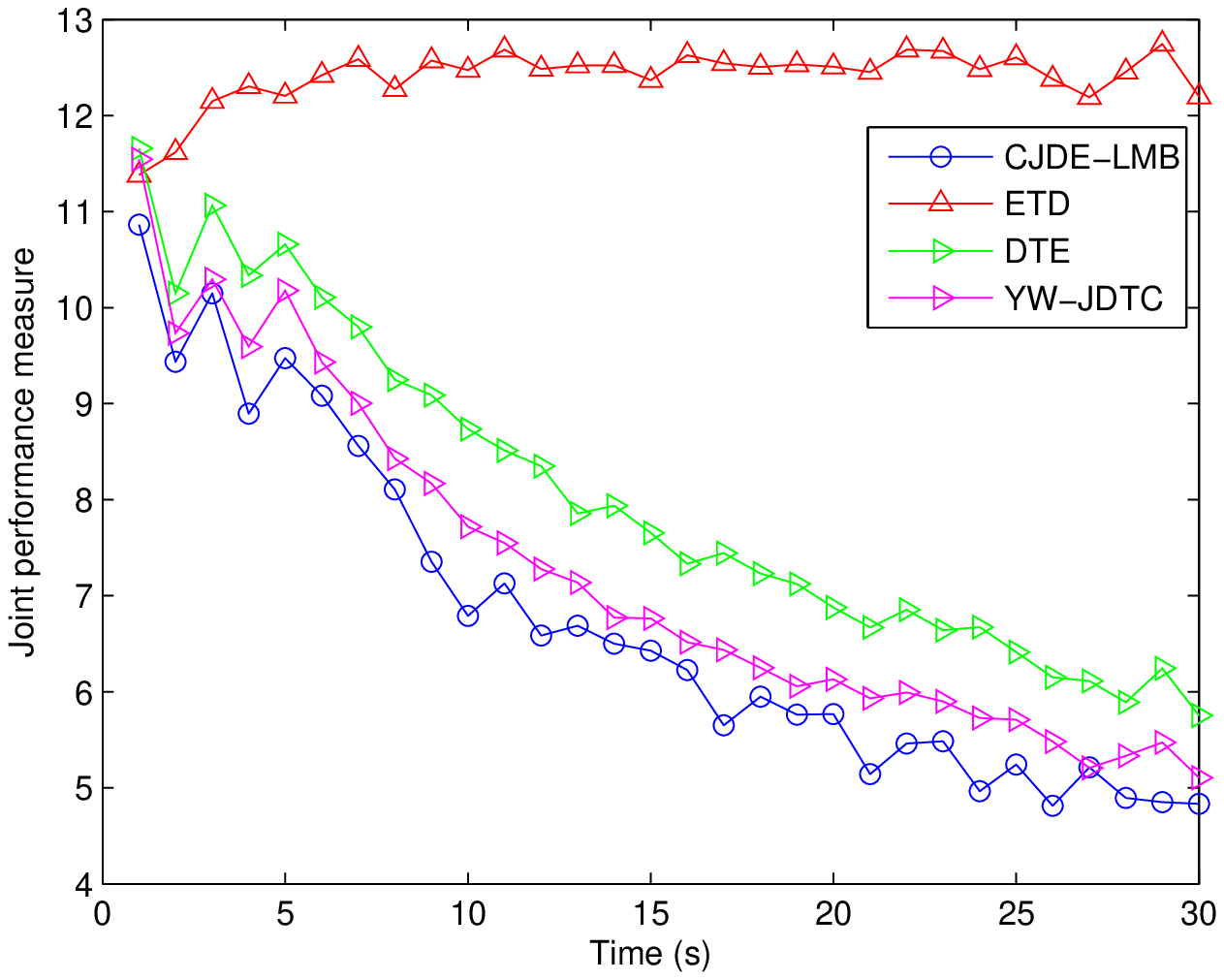}
	}
	\caption{The multi-target JDTC results. (a) Cardinality estimate, (b) OSPA distance, (c) Probability of incorrect Classification, (d) Joint performance metric.}
	\label{fig_images_org}
\end{figure*}

\indent Figure 1(a) illustrates the estimate of the multi-target cardinality. The targets are correctly detected by the proposed CJDE-LMB approach. The reason is that, because the coefficient $\gamma$ in the new CJDE risk is relatively large, the penalty of the target miss detection is severe. The tracking performance is shown in Fig. 1(b). As illustrated, the CJDE-LMB is the best in terms of the OSPA distance. The explanation of this result is that the interdependence between the decision and the estimation is considered, and the multi-target states are updated with reasonable MTA's. On contrary, the decision of the target class is not regarded in tracking when using ETD and YW-JDTC methods, and the error of the decision is not considered in the DTE method. Fig. 1(c) shows the classification results. The CJDE-LMB algorithm also performs best while the ETD method is the worst. The reason for this phenomenon is that the decision is only dependent on the current state estimation in the ETD method. In addition, although the superiority of the proposed CJDE-LMB algorithm over the YW-JDTC method is not very obvious, CJDE-LMB provides explicit decisions of the target classes, whereas YW-JDTC only computes the class probabilities. 
Summing up all the costs and the overall performance is evaluated in terms of the JPM. As depicted in Fig. 1(d), the performance of the CJDE-LMB algorithm is better than that of the other methods. 
This example shows that the performance of estimation and decision are improved because the interdependence between them are considered. Moreover, the proposed algorithm achieves the final goal directly and the explicit estimation and classification result are derived. \\

\subsection{Example 2}
\indent In order to illustrate the importance of the coefficients in the new Bayesian risk, the JDTC results are derived with different parameters in this example. Suppose that the coefficients are set to be $\alpha_{ij}^1=20, \beta_{ij}^1=1, \gamma_{ij}^1=100$, and $\alpha_{ij}^2=20, \beta_{ij}^2=1, \gamma_{ij}^2=10$, respectively. The values of $\alpha$ and $\beta$ make the costs of state estimation and classification balance. When $\gamma=100$, the target detection plays a dual role as before, on contrary, when $\gamma=10$, the cost of target miss detection contributes to $\bar{R}_C$ less significantly. \\
\begin{figure*}[!htp]
	\centering
	\subfigure[Cardinality estimate]{
		\label{fig_aerial_org}
		\includegraphics[width=2.5in]{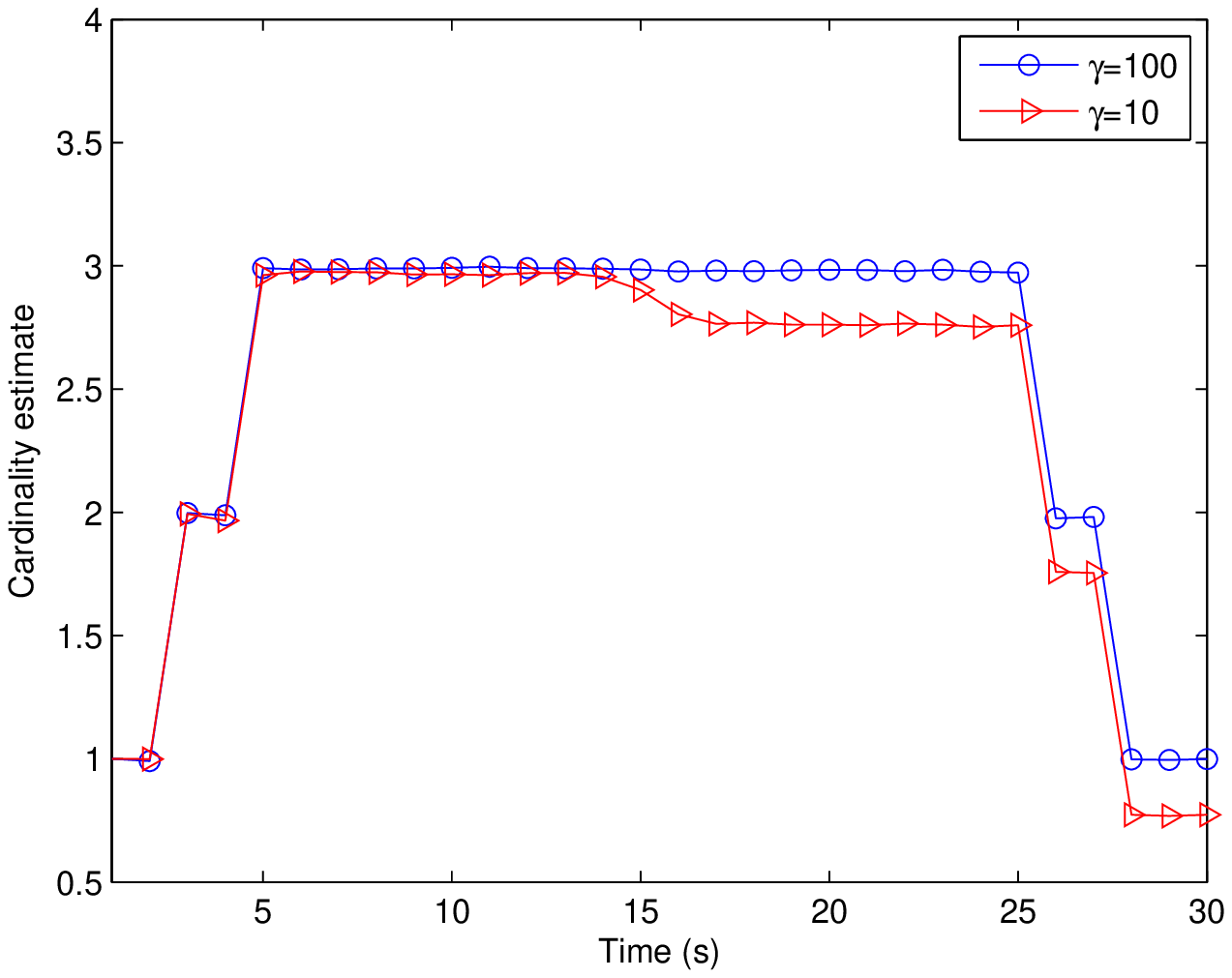}
	}
	\subfigure[OSPA distance]{
		\label{fig_barbara_org}
		\includegraphics[width=2.5in]{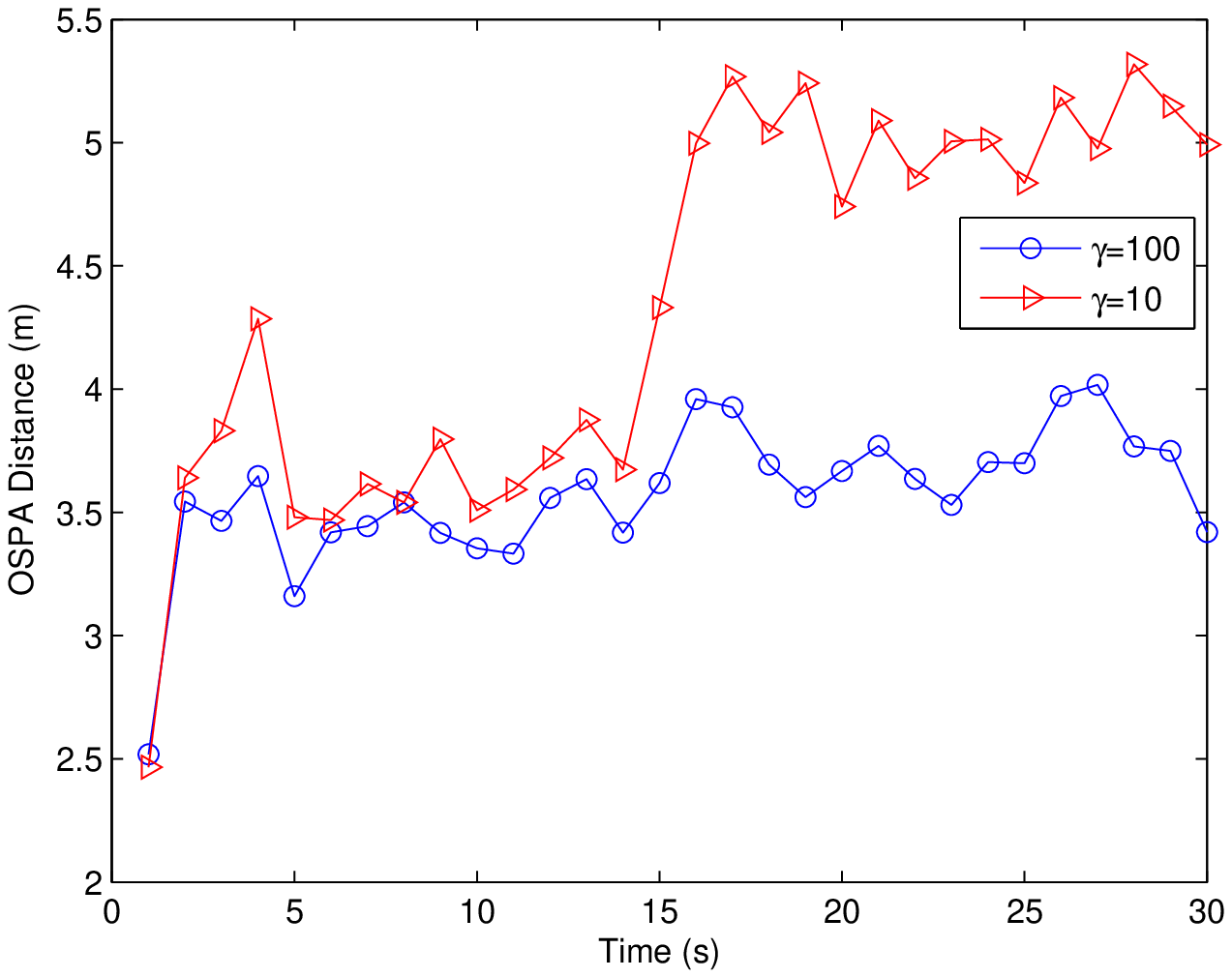}
	}
	\subfigure[Average probability of incorrect Classification]{
		\label{fig_boat_org}
		\includegraphics[width=2.5in]{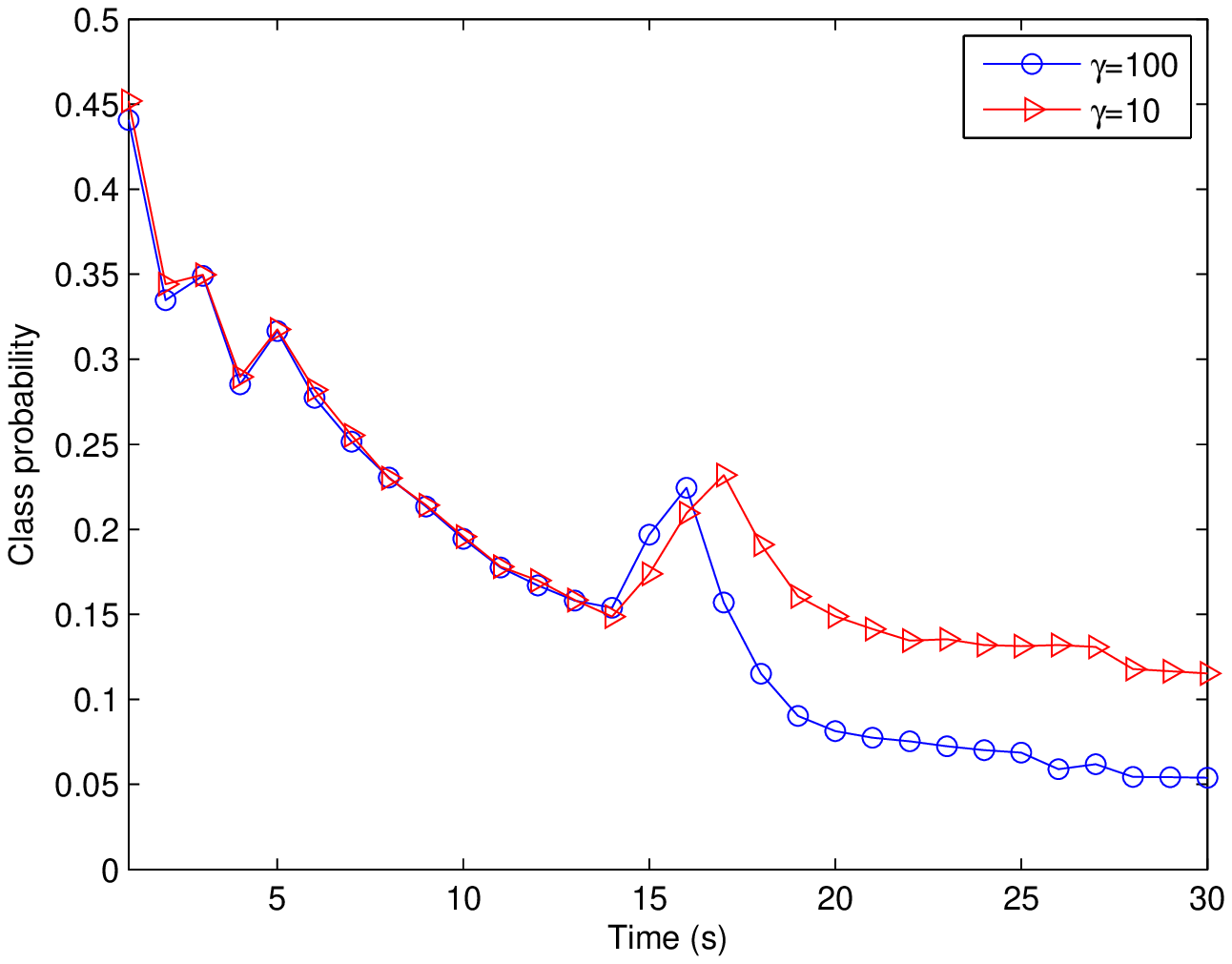}
	}
	\subfigure[Joint performance metric]{
		\label{fig_couple_org}
		\includegraphics[width=2.5in]{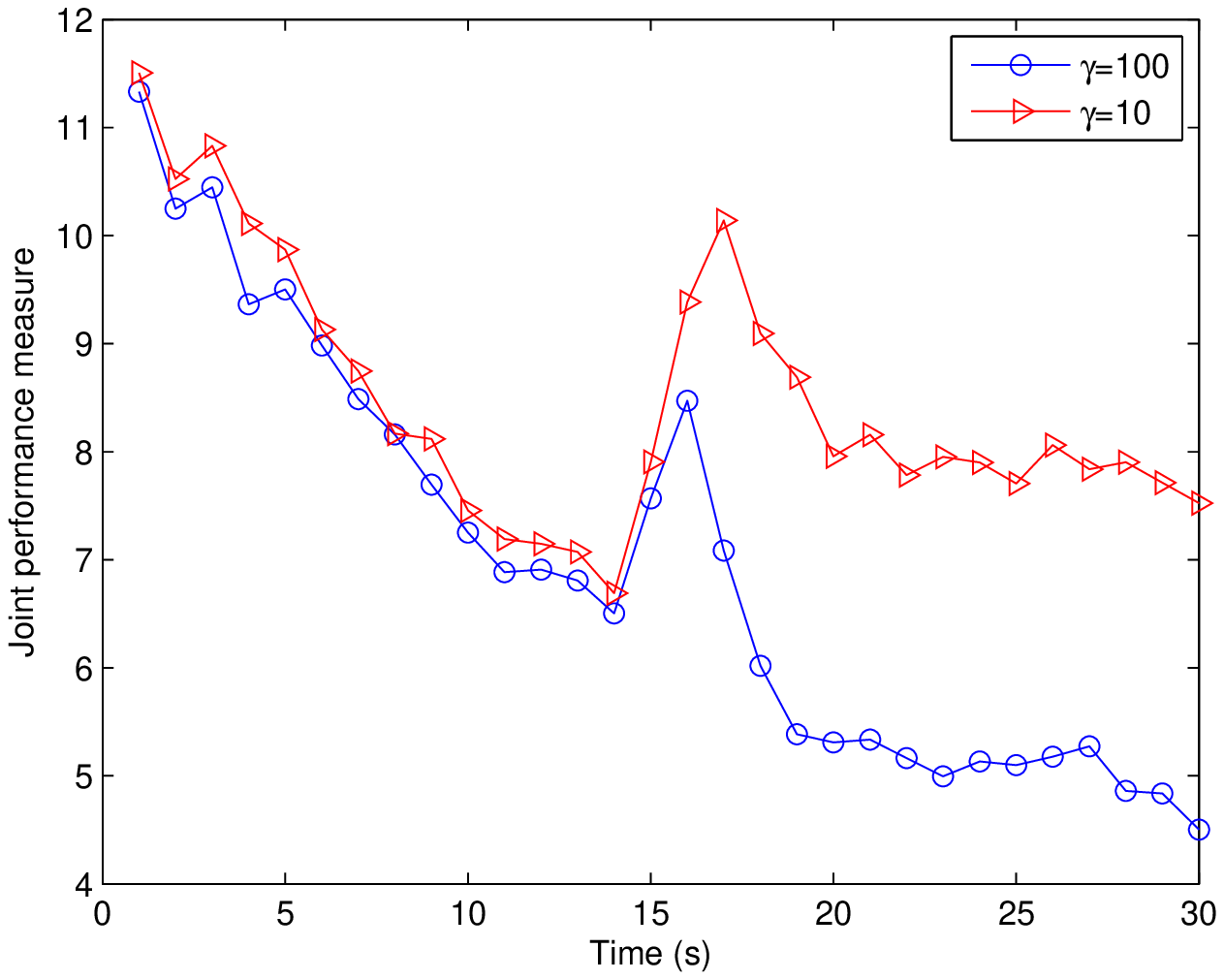}
	}
	\caption{The multi-target JDTC results. (a) Cardinality estimate, (b) OSPA distance, (c) Probability of incorrect Classification, (d) Joint performance metric.}
	\label{fig_images_org}
\end{figure*}
\indent The performance of target detection, tracking and classification under different parameters is illustrated in Fig. 2. As shown in Fig. 2(a), when all the targets keep their motion modes, all the tracks are detected correctly. After the target 3 executes constant acceleration, all the tracks are maintained under $\gamma=100$, whereas there exists target miss detection on some trials under $\gamma=10$. The reason for this phenomenon is that after the target 3 performs maneuver, the optimal Bayesian decision converts to maneuvering, both the costs of estimation and decision increase due to the transition of dynamic model and the change of optimal Bayesian decision, respectively. In this case, all the targets can be correctly detected when $\gamma=100$ because the penalization is heavier on target miss detection. On contrary, the decision with less state estimation and classification costs is chosen when $\gamma=10$, in this case, the target is judged to be undetected.
Due to the incorrect target detection results, the average tracking and classification performance given $\gamma=10$ is worse than $\gamma=100$ as illustrated in Fig. 2(b) and 2(c). As a result, the overall performance given $\gamma=100$ is also better as shown in Fig. 2(d). \\
\indent This example shows that, because target detection is the prerequisite for accurate tracking and correct classification in the multi-target JDTC problem, the penalization on target miss detection need to be heavier. 

\section{Conclusion}
In this paper, a novel recursive approach was proposed to solve the multi-target joint detection, tracking, and classification problem. 
The optimal solution was derived based on a new generalized Bayesian risk involving the costs of target number estimation, state estimation and classification. 
Because the interdependence between the decision and estimation was considered, the performances of multi-target detection, tracking and classification were improved. Moreover, as the multi-target density was approximated by a sum of class dependent components, the computational complexity was largely reduced.
The performance of the proposed approach was also analyzed, and the method of the coefficient selection was provided in order to derive reasonable results.
As illustrated in the simulations, the targets can be detected correctly under appropriate cost coefficients, and the state estimation and classification performances of the proposed approach were better than traditional methods. \\

\section*{Acknowledgment}
This work is jointly supported by National Natural Science Foundation of China (Grant Nos. 61673262 and 61175028), Shanghai key project of basic research (Grant No. 16JC1401100).

\end{document}